\documentclass[aps,pra,preprint]{revtex4}

\usepackage{amsmath}
\usepackage{amssymb}
\usepackage{bm}
\usepackage{graphicx}

\begin{document}

\title{Trapping time statistics and efficiency of
transport of optical excitations in dendrimers}

\author{Dirk-Jan Heijs}
\author{Victor A. Malyshev}
\author{Jasper Knoester}
  \email{knoester@phys.rug.nl}
\affiliation{%
Institute for Theoretical Physics and Materials Science Center,
University of Groningen, Nijenborgh 4,
9747 AG Groningen, The Netherlands
}%

\date{\today}

\begin{abstract}
We theoretically study the trapping time distribution and the
efficiency of the excitation energy transport in dendritic
systems. Trapping of excitations, created at the periphery of the
dendrimer, on a trap located at its core, is used as a probe of the
efficiency of the energy transport across the dendrimer. The
transport process is treated as incoherent hopping of excitations
between nearest-neighbor dendrimer units and is described using a
rate equation. We account for radiative and non-radiative decay of
the excitations while diffusing across the dendrimer. We derive exact
expressions for the Laplace transform of the trapping time distribution
and the efficiency of trapping and analyze those for various realizations
of the energy bias, number of dendrimer generations, and relative
rates for decay and hopping. We show that the essential parameter that
governs the trapping efficiency, is the product of the on-site
excitation decay rate and the trapping time (mean first passage time)
in the absence of decay.
\end{abstract}

\maketitle

\section{introduction}
\label{intro}

\begin{figure}[ht]
  \includegraphics[width=12cm]{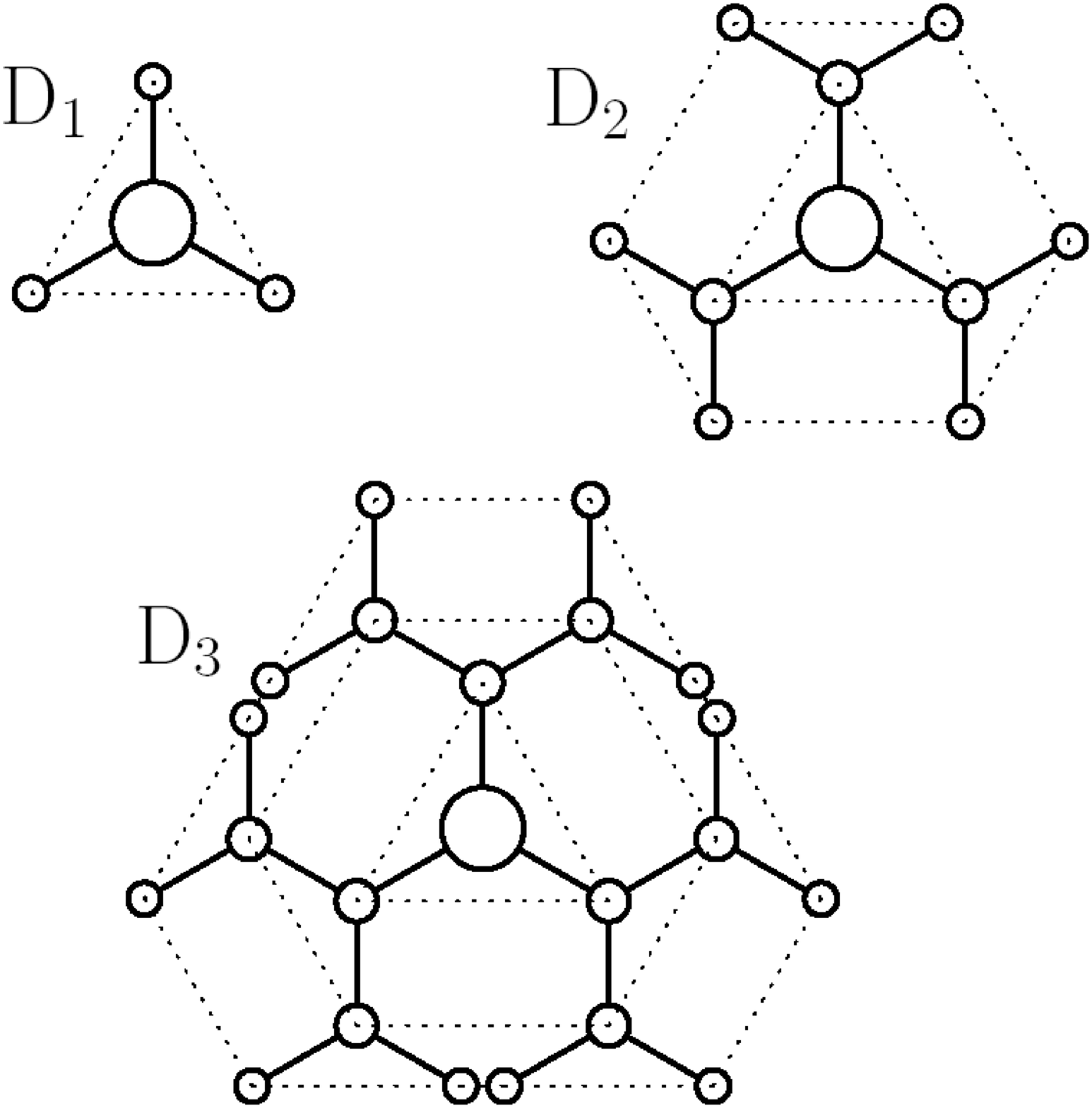}
  \caption{
    \label{fig:dendrimer}
    Schematic pictures of dendrimers of one (D$_1$), two
    (D$_2$), and three (D$_3$) generations with the coordination number
    $z = 3$ (equal to the number of branches at each branching point).
    The large circle at the dendrimer center represents a trap, while
    the small circles are building units of the dendrimer branches, representing
    sites on which optical excitations can reside.
    Connected by the dotted lines are the dendrimer units belonging to
    the same generation.
  }
\end{figure}

During the past decade dendritic molecular systems or dendrimers
have received considerable attention~\cite{Tomalia90,Frechet94,
Mukamel97,Jiang97,Vogte98,Frechet01}. Dendrimers are synthetic
highly branched tree-like macromolecules consisting of a core and
several branches which are not connected geometrically to each
other and are built self-similarly. Theoretically the process of
building a dendrimer can be repeated {\em ad infinitum} to obtain a
dendrimer with any number of generations, but in practice this
number is currently limited to fifteen~\cite{Frechet01}. The first
three dendrimers with coordination number $z = 3$ are schematically
depicted in Fig.~\ref{fig:dendrimer}.

Dendrimers hold great promise for creating artificial light
harvesting systems. Indeed, because of their branched nature, the
number of units at their periphery grows exponentially with the
number of generations . Therefore, if light absorbing states are
located at the periphery, the cross-section of absorption grows
exponentially with the number of generations. Combining this with
a possibly efficient energy transport to the core may result in
efficient light harvesting
systems~\cite{Frechet94,Mukamel97,Jiang97}.

A growing flux of publications exists, both
experimental~\cite{Devadoss96,Kopelman97,Jiang98,Hofkens99,Yeow00,%
Varnavski00,Andronov00,Drobizhev00,Texier01,Kleiman01,Varnavski02}
and
theoretical~\cite{Barhaim97,Barhaim98,Rana03,Tretiak98,Raychaudhuri00,%
Argyrakis00,Poliakov99,Bentz03,Nakano00,Kirkwood01,Reineker01,Nakano01},
reporting on optical and transport properties of dendritic
systems. For instance, in Ref.~\onlinecite{Kopelman97} compact and
extended polyphenylacetylene dendrimers in solution were studied
experimentally. Interpreting their spectroscopic measurements, the
authors argued that the optical excitations in these systems are
localized on dendrimer subunits. They also made the important
point that extended dendrimers were characterized by a so-called
energy funnel: the excitation energies of dendrimer units
decreases from the periphery towards the core, thus providing an
energetic bias for transport of the excitation towards the core.
Such a bias is favorable for efficient transport of the absorbed
energy across the dendrimer. Using quantum chemical calculations,
the existence of an energy funnel towards the core was further
confirmed in Ref.~\onlinecite{Tretiak98}. Experimental
investigations of the energy transport in these
systems~\cite{Devadoss96} revealed the fact that it occurred
through a multi-step (incoherent) hopping process with an
efficiency of 96\%. On the other hand, low-temperature measurements
of the energy transport in distyrylbenzene-stilbene dendrimers
with a nitrogen core gave clear indications of coherent
interactions between the dendrimer subunits~\cite{Varnavski02}. In
this class of organic dendrimers, ultra-fast higher-order
nonlinearities were also reported~\cite{Varnavski00}, which makes
them potentially promising systems for applications in nonlinear
optical switching elements.

The first theoretical efforts on dendrimers were mostly focused on
analyzing the mean first passage time, i.e., the average time it
takes for an excitation, created somewhere at the periphery of the
dendrimer, to reach its center, where a trap is located.  This
problem was addressed in much detail in
Refs.~\cite{Barhaim97,Barhaim98,Rana03}, where the mean first
passage time was calculated both in the presence and the absence of
a fixed energy bias. The presence of a random bias was considered
in Ref.~\cite{Raychaudhuri00}, where it was found that the
randomness tends to reduce the transport efficiency. Large-scale
numerical calculations of the statistics of the number of sites
visited by the excitation and its mean square displacement in very
large dendritic structures were performed in
Ref.~\cite{Argyrakis00}, while the authors of Ref.~\cite{Bentz03}
discussed the kinetics of symmetric random walks in compact and
extended dendrimers of a small number of generations ($\le 4$).

In all papers cited above, the basis of the analysis was the
assumption of incoherent hopping transport between dendrimer
units. However, the actual nature of the excitation transport and
optical dynamics is still under debate. Thus, the alternative
approach of coherent excitons extending over several dendrimer
units has also been considered, in particular to describe energy
transport from the periphery to the
core~\cite{Poliakov99,Nakano00,Kirkwood01,Reineker01} and the
enhancement of the third-order optical
susceptibility~\cite{Nakano01}. An other interesting branch of
experimental and theoretical studies of dendritic systems concerns
the dynamics of dendrimer-based networks, i.e., extended systems
using dendrimers as building
blocks~\cite{Jahromi01,Gitsov02,Dvornic02,Gurtovenko03}. Recently,
such networks have attracted much attention due to their two levels
of structural organization.

Among the factors that govern the efficiency of energy transport
in dendritic systems, such as the energy bias and the presence of
disorder, the radiative and non-radiative decay of optical
excitations during their random walk to the core represents an
additional channel, especially for dendrimers with a large number
of generations. To the best of our knowledge, this factor has thus
far not been addressed in the literature. Moreover, the statistics
of the trapping time has not been studied in detail either. In this
paper, we intend to fill these gaps and show that intimate
relations exist between the analysis of both issues. We will show
that in the presence of decay, the key parameter that governs the
trapping efficiency is the product of the on-site excitation decay
rate and the mean first passage time in the absence of decay.

The outline of this paper is as follows. In Section~\ref{model} we
present our model, which is based on incoherent motion of
excitations across the dendrimer, described by a rate equation that
accounts for both trapping at the core and excitation decay during
the random walk towards the core. We also discuss several
quantities relevant to the problem under study, such as the
distribution of survival times, the mean survival time, and the
efficiency of trapping. Section~\ref{STDexact} deals with deriving
exact expressions for these quantities in the Laplace domain. In
Sec.~\ref{STDproperties}, we present a detailed analysis of the
trapping time distribution in the limit of vanishing excitation
decay. Effects of the excitation decay on the trapping efficiency
under various conditions (sign of the energy bias, number of
generations, ratio of decay and hopping rates) are discussed in
Sec.~\ref{ET}. Finally, we conclude in Sec. ~\ref{concl}.

\section{Model and pertinent quantities}
\label{model}

As was already mentioned in the Introduction, we treat both the
motion of the excitation over the dendrimer units and the trapping
at the core as incoherent nearest-neighbor hopping processes. We
label the dendrimer units (sites) by the index $i$ ($1 \le i \le
\cal{N}$), where $\cal{N}$ is the number of units, while $i = 0$
denotes the trap located at the core (cf.~Fig.~1). The trap is
considered irreversible, i.e., once the excitation hops onto it,
it never returns to the body of dendrimer. Then, the system of rate
equations for the excitation probabilities (populations) $p_i$ of
the dendrimer units reads
\begin{subequations}
\label{master_equation}
\begin{equation}
\label{p0}
    {\dot p}_0 = \sum_{\{j\}} k_{0j}p_j \ ,
\end{equation}
\begin{equation}
\label{pi}
    {\dot p}_i = -\gamma p_i
    -\sum_{\{j\}}k_{ji}p_i +{\sum_{\{j\}}}^\prime k_{ij}p_j \ \quad (i\ne 0).
\end{equation}
\end{subequations}
Here, the dot denotes the time derivative, the summation
$\sum_{\{j\}}$ is performed over sites $j$ that are
nearest-neighbors of the site $i$, the prime in the second
summation of Eq.~(\ref{pi}) indicates that $j \ne 0$, $\gamma$ is
the exciton decay rate (assumed independent of $i$), and $k_{ij}$
is the rate of hopping from site $j$ to site $i$, including $i =
0$. The hopping rates meet the principle of detailed balance:
$k_{ij} = k_{ji} \exp[(E_j-E_i)/k_B T]$, where $E_i$ is the
excitation energy of the site $i$, $k_B$ is the Boltzmann constant,
and $T$ is the temperature. In this sense the energy of the trap,
$E_0$, is considered infinitely low. Initially, the excitation is
outside the trap, i.e., $p_0(0) = 0$, while one of dendrimer units
is excited, $p_i(0) = \delta_{ii_0}$.

The quantity $r(t) \equiv {\dot p}_0$ represents the instantaneous
trapping rate and will be used to study the time-domain behavior
of the energy transport in dendritic systems. It can be expressed
through the total population outside the trap, $\psi =
{\sum}^\prime_i p_i$, which is hereafter referred to as the
survival probability with respect to both decay and trapping.
Indeed, from Eqs.~(\ref{master_equation}) it follows that $\sum_i
{\dot p}_i \equiv {\dot p}_0 + {\dot \psi} = -\gamma \psi$, so
that for $r(t)$ one finds
\begin{equation}
\label{r} r(t) =  -\gamma \psi - {\dot \psi} \ .
\end{equation}
Furthermore, the time dependence of $r(t)$ deriving from the decay
constant $\gamma$, can be extracted explicitly by the
transformation $p_i = e^{-\gamma t}\tilde{p}_i$. After this
transformation Eq.~(\ref{r}) is reduced to
\begin{equation}
\label{R}
    r(t) = - e^{-\gamma t} {\dot \Psi} \equiv e^{-\gamma t}R(t) \ ,
\end{equation}
where $\Psi(t) = {\sum}^\prime_i \tilde{p}_i$, and the
$\tilde{p}_i$ now obey Eq.~(\ref{pi}) with $\gamma = 0$. Thus,
$\Psi$ is the analog of $\psi(t)$ and represents the total
population outside the trap in the absence of excitation decay. It
is the survival probability with respect to trapping alone. We see
from Eq.~(\ref{R}) that in the time domain, the trapping and the
decay of the excitations are independent of each other. Therefore,
the time behavior of the trapping process can be studied
separately, which simplifies the analysis.

The quantity $R(t) \equiv -{\dot\Psi}$ is normalized to unity for
finite systems (in sense that $\int_0^\infty R(t)dt = 1$) and
represents the probability distribution of the (pure) trapping
time. From this, the mean trapping time, often referred to as the
mean first passage time~\cite{Barhaim97}, is calculated in a
standard way
\begin{equation}\label{mean_t_Psi}
    \langle t\rangle =
    \int_{0}^{\infty}t R(t)\,{\rm d}t = \int_{0}^{\infty}
    \Psi(t)\,{\rm d}t \ .
\end{equation}
The inverse quantity $\langle t \rangle^{-1}$ represents the
effective trapping rate in the absence of excitation decay. Note
that Eq.~(\ref{mean_t_Psi}) can be rewritten via the Laplace
transform $\tilde{R}(s) = \int_{0}^{\infty}e^{-st}R(t)\,{\rm d}t$
of $R(t)$:
\begin{eqnarray}\label{mean_t_Laplace}
    \langle t\rangle =
    -\frac{{\rm d}\tilde{R}}{{\rm d}s}\Big|_{s=0} \ ,
\end{eqnarray}
which is useful for further considerations (see below).

We now define the efficiency of trapping (denoted as
$\varepsilon$) as the total population that is transferred to the
trap, i.e., the fraction of the initially created excitation that
reaches the trap during its lifetime:
\begin{equation}
\label{q}
    \varepsilon \equiv \lim_{t\rightarrow\infty}p_0(t)
    = \int_{0}^{\infty}r(t)\,{\rm d}t
    = \int_{0}^{\infty}e^{-\gamma t}R(t)\,{\rm d}t
    = \tilde{R}(s)|_{s=\gamma}\ .
\end{equation}
An other important quantity that contains information about the
trapping efficiency is
\begin{equation}
\label{tau}
    \tau = \int_{0}^{\infty}\psi(t)\,{\rm d}t =
    \frac{1-\varepsilon}{\gamma} \ ,
\end{equation}
which is the mean survival time (with respect to both trapping and
decay). Using this definition, we can, by convention, define the
effective trapping rate $W$ in the presence of excitation decay as
\begin{equation}
\label{W}
    W = \frac{1}{\tau} - \gamma
    = \frac{\varepsilon}{1-\varepsilon}\gamma \ ,
\end{equation}

It should be stressed that $\varepsilon$, $\tau$, and $W$, being
defined through time-integrations, are influenced by the excitation
decay. In particular, if the latter occurs on a time scale much
slower than the mean first passage time $\langle t \rangle$, the
trapping efficiency $\varepsilon$ is close to unity, the survival
time $\tau$ is reduced to the mean first passage time $\langle
t\rangle$, given by Eq.~(\ref{mean_t_Psi}), and $W \approx \langle
t\rangle^{-1}$. If however the excitation decay occurs on a time
scale that is comparable to or faster than the mean first passage
time, the quantities $\varepsilon$, $\tau$, and $W$ will be
determined by the interplay of the random walk to the trap and the
excitation decay. This interplay between trapping and decay will
be one of the main issues in the remainder of this paper.

To conclude this section we notice that all quantities relevant to
the excitation energy transport, such as $\langle t\rangle$,
$\varepsilon$, $\tau$, and $W$, are directly related to the Laplace
transform $\tilde{R}(s)$ of the trapping time distribution $R(t)$
in the absence of decay. In the next section, we will provide the
exact solution for $\tilde{R}(s)$.

\section{Laplace domain analysis: exact results}
\label{STDexact}

\begin{figure}[ht]
  \includegraphics[width=12cm]{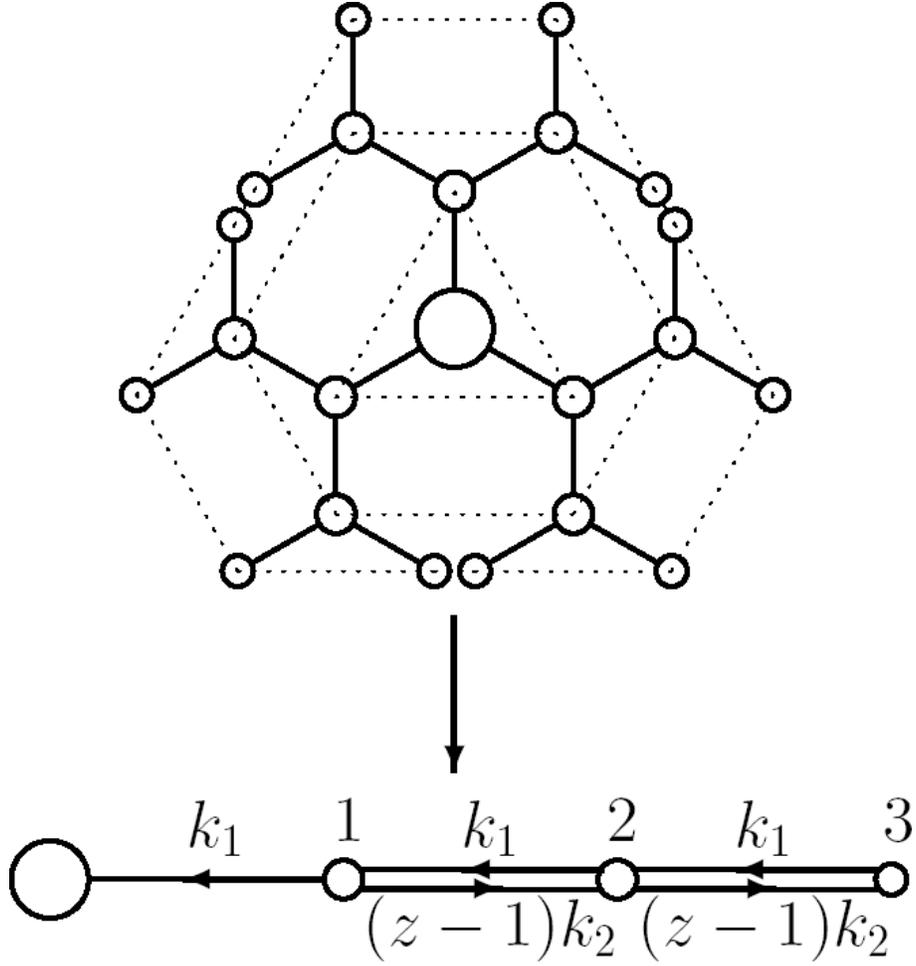}
  \caption{
    \label{fig:mapping}
    Example of mapping of a dendrimer D$_3$ with $z = 3$
    and a linear energy bias onto an equivalent linear chain. The
    large circle represents the trap, while all units of generation $M
    (=1,2,3)$ are mapped onto one site $M$ of the linear chain, which is
    drawn by a small circle. The quantities $k_1$ and $k_2$ are the
    hopping rates towards and away from the trap, respectively, in the real
    dendrimer. The factor $z-1$ counts the number of branches towards the
    periphery at each branching point and multiplies $k_2$ to obtain the
    effective outward hopping rate in the equivalent linear chain.
  }
\end{figure}

From now on, we will consider a specific model for the hopping
process, in which only two different hopping rates occur.
Specifically, we will assume that the hopping rates towards and
away from the dendrimer's core are, respectively, $k_1$ and $k_2$,
no matter at which branching point of the dendrimer the excitation
resides. This assumption corresponds to the situation with a
linear energy bias, where the excitation energy difference between
units of generation $M$ and $M-1$ is a constant, $\Delta E$, which
is identical for every $M$. As was pointed out in
Ref.~\cite{Hughes82}, in this case the random walk across the
dendrimer may be mapped onto a random walk on an asymmetric linear
chain, where the rate of hopping is $k_1$ in the direction of one
end of the chain (where the trap resides) and $(z-1)k_2$ in the
other direction. In other words, instead of a dendrimer of
generation $N$ a linear chain of length $N+1$ is considered with a
trap at site $0$. This mapping is illustrated in
Fig.~\ref{fig:mapping}.

After this mapping, the set of rate equations is different for
dendrimers of one, two, and $N > 2$ generations. For a
one-generation dendrimer, only one equation occurs:
\begin{equation}
\label{D1}
    \dot{P}_1 = -k_1 P_1 \ .
\end{equation}
For a dendrimer with two or more generations, we have the
following equations:
\begin{subequations}
\label{DN}
\begin{equation}
\label{DN1}
    \dot{P}_1 = -\left[k_1 + (z-1)k_2\right]P_1 + k_1P_2 \ ,
\end{equation}
\begin{equation}
\label{DN2}
    \dot{P}_M = (z-1)k_2P_{M-1}-\left[k_1 + (z-1)k_2\right] P_M
    + k_1P_{M+1} \ , \quad 1 < M < N \ ,
\end{equation}
\begin{equation}
\label{DN3}
    \dot{P}_N = (z-1)k_2 P_{N-1} - k_1P_N \ ,
\end{equation}
\end{subequations}
where in case $N=2$, Eq.~(\ref{DN2}) is absent. In all these
equations, $P_M$ denotes the total population in the $M$th
dendrimer generation, while the factor $z-1$ accounts for the
number of nearest-neighbor units towards the periphery for each
branching point. From this form it is clearly seen that the
branching ($z \ge 2$) leads to a ``geometrical" bias towards the
dendrimer's periphery, even in the absence of an energetic bias
($\Delta E = 0$). Whether a net bias exists and in what direction,
depends on the quantity $\kappa=(z-1)k_2/k_1 = (z-1)\exp({-\Delta
E}/k_BT)$. At $\kappa=1$ ($k_1 = (z - 1)k_2$), the geometrical and
energetic biases exactly compensate each other, while for $\kappa
< 1$ ($\kappa>1$) a net bias towards (away) from the trap occurs.

As initial condition we will consider the situation where one
excitation has been created at the periphery of the dendrimer,
i.e., $P_M(0) = \delta_{MN}$. According to Eq.~(\ref{p0}), the
trapping rate in the absence of excitation decay is now given by
$R(t) = k_1P_1(t)$. This will be the quantity of our prime
interest, for which we will seek a solution in the remainder of
this section. Solving the one- and two-generation dendrimer
problems is straightforward and will be done later on. Our main
goal is to find the solution of the general problem of an
$N$-generation dendrimer. This may be done in the Laplace domain.
If for brevity we introduce the dimensionless time $t^\prime =
k_1t$, Eqs.~(\ref{DN}) written in the Laplace domain take the form:
\begin{subequations} \label{laplace}
\begin{equation}
\label{P1laplace1}
    0 = -\left(1+\kappa+s\right)\tilde{P}_1+\tilde{P}_{2} \ ,
\end{equation}
\begin{equation}
\label{PMlaplace1}
     0 = \kappa \tilde{P}_{M-1}-\left(1+\kappa+s\right)\tilde{P}_M
     +\tilde{P}_{M+1} , \ \quad 1 < M < N \ ,
     \end{equation}
\begin{equation}
\label{PNlaplace1}
    -1 = \kappa \tilde{P}_{N-1}-\left(1+s\right)\tilde{P}_N \ ,
\end{equation}
\end{subequations}
where the Laplace parameter $s$ is now in units of $k_1$ and
$\tilde{R} = \tilde{P_1}$ (we use a tilde to denote the Laplace
transformed functions). Below, we find a recursive relation for
$\tilde{R}$, connecting this quantity for dendrimers of different
numbers of generations (i.e., lengths of the effective linear
chain). Therefore, we will from now on denote $\tilde{R}$ for a
dendrimer of $N$ generations as $\tilde{R}_N$.

After $N-2$ steps of eliminating $\tilde{P}_N, \tilde{P}_{N-1},...,
\tilde{P}_3$ from Eqs.~(\ref{PMlaplace1}) and~(\ref{PNlaplace1}),
we arrive at two coupled equations
\begin{subequations}
\begin{equation}
\label{P1laplace2} 0 =
-\left(1+\kappa+s\right)\tilde{P}_1+\tilde{P}_2 \ ,
\end{equation}
\begin{equation}
\label{P2laplace2}
    -B_N = A_N \tilde{P}_1-\tilde{P}_2 \ ,
\end{equation}
\end{subequations}
a solution of which with respect to $\tilde{P}_1 = \tilde{R}_N$ is
\begin{equation}
\label{RN}
    \tilde{R}_N = \frac{B_N}{1+\kappa+s-A_N} \ .
\end{equation}
Here, $A_N$ and $B_N$ are functions of the Laplace parameter $s$,
which will be specified later on. For a dendrimer of $N+1$
generations, similarly, $N-2$ steps of excluding $\tilde{P}_{N+1},
\tilde{P}_{N},..., \tilde{P}_4$ from Eqs.~(\ref{PMlaplace1})
and~(\ref{PNlaplace1}) yield a system of three coupled equations
\begin{subequations}
\begin{equation}
\label{P1laplace3}
    0 = -\left(1+\kappa+s\right)\tilde{P}_1+\tilde{P}_2 \ ,
\end{equation}
\begin{equation}
\label{P2laplace3}
    0 = \kappa \tilde{P}_{1}-\left(1+\kappa+s\right)\tilde{P}_2
    + \tilde{P}_3 \ ,
\end{equation}
\begin{equation}
\label{P3laplace3}
    -B_N = A_N \tilde{P}_2-\tilde{P}_3 \ ,
\end{equation}
\end{subequations}
where $A_N$ and $B_N$ are the same as in
Eqs.~(\ref{P1laplace2}),~(\ref{P2laplace2}), and~(\ref{RN}), while
now $\tilde{P}_1 = \tilde{R}_{N+1}$. Solving
Eqs.~(\ref{P1laplace3})-(\ref{P3laplace3}) with respect to
$\tilde{P}_1$ leads to an expression that is algebraically
identical to Eq.~(\ref{RN}), except that $A_N$ and $B_N$ are
replaced by $A_{N+1}$ and $B_{N+1}$, respectively. The latter are
given by
\begin{subequations}
\label{ABrecursive}
\begin{eqnarray}
A_{N+1} = \frac{\kappa}{1+\kappa+s-A_N} \ ,
\end{eqnarray}
\begin{eqnarray}
B_{N+1} = \frac{B_N}{1+\kappa+s-A_N} \ ,
\end{eqnarray}
\end{subequations}
and represent the recursive relations for these two functions.
From comparison of Eq.~(\ref{ABrecursive}) with Eq.~(\ref{RN}),
one finds that $B_{N+1} = \tilde{R}_N$ and $A_{N+1} =
\kappa\tilde{R}_N/\tilde{R}_{N-1}$. Substituting these relations
back into Eq.~(\ref{RN}), we arrive at a recursive relation for
$\tilde{R}_N$:
\begin{equation}
\label{recursive_relation}
    \frac{1}{\tilde{R}_N} =
    (1+\kappa+s)\frac{1}{\tilde{R}_{N-1}}
    -\kappa \frac{1}{\tilde{R}_{N-2}} \ .
\end{equation}

Note that Eq.~(\ref{recursive_relation}) can be used for $N > 2$
only. In order to close the recursive iteration, $\tilde{R}_1$ and
$\tilde{R}_2$ must be calculated separately. In the Laplace domain
the solution of Eq.~(\ref{D1}) for $N=1$ and Eqs.~(\ref{DN1}) and
(\ref{DN3}) for $N=2$ is straightforward and gives us the necessary
quantities:
\begin{subequations}
\begin{equation}
\label{R1}
    \tilde{R}_1 =  \frac{1}{1+s} \ ,
\end{equation}
\begin{equation}
\label{R2}
    \tilde{R}_2 = \frac{1}{1 + (2 + \kappa) s + s^2} \ .
\end{equation}
\end{subequations}
Using Eqs.~(\ref{R1}) and (\ref{R2}) in the recursive relation
Eq.~(\ref{recursive_relation}) and solving this relation, we
finally obtain $\tilde{R}_N(s)$ in closed form:
\begin{eqnarray}
\label{closedRN}
    \frac{2^{N+1}}{\tilde{R}_N} &=& \left(
    1+\frac{1-\kappa+s}{\sqrt{(1+\kappa+s)^2-4\kappa}} \right)
    \left(1+\kappa+s+\sqrt{(1+\kappa+s)^2-4\kappa}\right)^N
\nonumber\\
\nonumber\\
&+& \left( 1-\frac{1-\kappa+s}{\sqrt{(1+\kappa+s)^2-4\kappa}}
\right) \left(1+\kappa+s-\sqrt{(1+\kappa+s)^2-4\kappa}\right)^N.
\end{eqnarray}
This is the principal result of this section.

It should be noticed that, despite the fact that
Eq.~(\ref{closedRN}) contains square roots, ${\tilde{R}_N}^{-1}$
in reality is a polynomial of $N$th order in $s$. This may be
checked by explicit expansion of Eq.~(\ref{closedRN}) in powers of
$s$. Alternatively, the recursive relation
Eq.~(\ref{recursive_relation}) itself already represents a proof of
this statement.

\section{Trapping time distribution}
\label{STDproperties}

It turns out to be impossible to perform the transformation back
to the time domain for the general result Eq.~(\ref{closedRN}),
implying that we do not obtain an explicit expression for the
trapping time distribution. However, the moments of this
distribution and its asymptotic form at large $N$ do allow for an
analytical treatment. In this section, we will address these two
topics. We will also show that the asymptotic form may already be
reached for rather small dendrimers, making these analytical
expressions of practical use.

\subsection{Moments}
\label{momenta}

To study the moments of the trapping time distribution, we make
use of the following relation:
\begin{equation}
\label{definition}
    \langle t^n\rangle_N \equiv \int_{0}^{\infty}t^n R_N(t)
    \,{\rm d}t =
    (-1)^n\frac{{\rm d}^n \tilde{R}_N}{{\rm d}s^n}\Big|_{s=0}\ .
\end{equation}
Furthermore, we recall that $1/\tilde{R}_N$ is a polynomial of
$N$th order in $s$ (see the previous section), and thus can be
written as
\begin{equation}
\label{polynomial}
    \frac{1}{\tilde{R}_N} = \sum_{m=0}^{N}a_N^{(m)}s^m \ ,
    \quad\quad
    a_N^{(m)} = \frac{1}{m!}\left(\frac{{\rm
    d}^m}{{\rm d}s^m} \frac{1}{\tilde{R}_N}\right)\Big|_{s=0}\ .
\end{equation}
Substituting Eq.~(\ref{polynomial}) into
Eq.~(\ref{recursive_relation}) and comparing coefficients related
to the same power of $s$, we obtain a recursive relation for
$a_N^{(m)}$
\begin{equation}
\label{recursive_relation_2}
    a_{N}^{(m)} = a_{N-1}^{(m-1)}
    +\left(1+\kappa\right)a_{N-1}^{(m)}-\kappa a_{N-2}^{(m)}\ ,
\end{equation}
where it is implied that $a_{N}^{(m)} = 0$ for $m
> N$ and $m < 0$. From Eqs.~(\ref{R1}) and (\ref{R2}) it follows
that for $N = 1$ and $N = 2$,
\begin{equation}
a_1^{(0)} = a_1^{(1)} = a_2^{(0)} = a_2^{(2)} = 1 \ , \quad
a_2^{(1)} = 2+\kappa \ .
\end{equation}
This allows us to start the recursive procedure for $a_N^{(m)}$.

The first important conclusion concerning the expansion
coefficients, which follows from Eq.~(\ref{recursive_relation_2}),
is that $a_N^{(0)} = 1$ for any $N$. This simply means that the
zero'th moment of the trapping time distribution $\tilde{R}_N(0) =
1$, i.e., the distribution is normalized. As a consequence of this
fact, one can elucidate the physical meaning of the second
coefficient of the expansion Eq.~(\ref{polynomial}). It appears to
be equal to the first moment of the distribution $R_N$, which is
nothing but the mean trapping time $\langle t\rangle_N$:
\begin{equation}
    a_N^{(1)} = \frac{\rm d}{{\rm d}s}\frac{1}{\tilde{R}_N}\Big|_{s=0}
    = -\left(\frac{1}{\tilde{R}_N^2}
    \frac{{\rm d}\tilde{R}_N}{{\rm d}s}\right)\Big|_{s=0}
    = \langle t\rangle_N \ .
\end{equation}
Substituting this result in Eq.~(\ref{recursive_relation_2}),
yields a recursive formula for $\langle t\rangle_N$
\begin{equation}
    \langle t\rangle_N = 1+(1+\kappa)\langle
t\rangle_{N-1}-\kappa\langle t\rangle_{N-2} \ ,
\end{equation}
from which earlier results for the mean trapping time in
dendrimers are reproduced~\cite{Barhaim97}:
\begin{subequations}
\label{MFPT}
\begin{eqnarray}
\label{keq1}
    \langle t\rangle_N
    = \frac{1}{2}N(N+1)\ ,
    \quad\quad\quad
    \kappa = 1 \ ,
\end{eqnarray}
\begin{eqnarray}
\label{kne1}
    \langle t\rangle_N = \frac{N}{1-\kappa}
    + \frac{\kappa}{(1-\kappa)^2}\left[\kappa^N-1\right] \ ,
    \quad\quad
    \kappa \ne 1 \ .
\end{eqnarray}
\end{subequations}

The next coefficient, $a_N^{(2)}$, contains information about the
second moment of $R_N$. It is straightforward to show that
\begin{equation}
\label{a2vsMomenta}
    a_N^{(2)}
    = \langle t\rangle_N^2-\frac{1}{2}\langle t^2\rangle_N \ .
\end{equation}
On the other hand, using Eq.~(\ref{recursive_relation_2}) and the
method of induction, one may show that
\begin{subequations}
\label{a_2}
\begin{equation}
    a_N^{(2)} = \frac{1}{24}(N-1)N(N+1)(N+2) \ ,
    \quad\quad \kappa = 1 \ ,
\end{equation}
\begin{eqnarray}
    a_N^{(2)} &=&
    \frac{\kappa^{N+1}[\kappa(N-1)-(N+2)]}{(1-\kappa)^4}
    \nonumber\\
    \nonumber\\
    &+& \, \frac{N(N+1)}{2(1-\kappa)^2}
    + \, \frac{\kappa^2(N+1)+2\kappa-N}{(1-\kappa)^4} \ ,
    \quad\quad\quad \kappa \ne 1 \ .
\end{eqnarray}
\end{subequations}
Making now use of Eqs.~(\ref{MFPT}) and (\ref{a_2}) in
Eq.~(\ref{a2vsMomenta}), we find $\langle t^2\rangle_N$:
\begin{subequations}
\begin{equation}
\label{t2keq1}
    \langle t^2\rangle_N = \frac{1}{12}N(N+1)[5N(N+1)+2] \ ,
    \quad\quad \kappa = 1 \ ,
\end{equation}
\begin{eqnarray}
    \langle t^2\rangle_N &=& \frac{N(N+1)}{(1-\kappa)^2} +
    \frac{(6N+2)\kappa^{N+1}}{(1-\kappa)^3}
    \nonumber\\
    \nonumber\\
    &+& \frac{2\kappa}{(1-\kappa)^4}[\kappa^{2N+1}+\kappa^{N}-2]
    \ , \quad\quad \kappa \ne 1 \ .
\end{eqnarray}
\end{subequations}

It is worthwhile to consider the asymptotic behavior of the first
and second moments of $R_N$ at large $N$, the case of our primary
interest. They are given by
\begin{subequations}
\label{limit_coeff}
\begin{eqnarray}
\label{1}
    \langle t\rangle_N = \frac{N}{1-\kappa} \ , \quad
    \langle t^2\rangle_N = \frac{N(N + 1)}{(1-\kappa)^2} \ ,
    \quad\quad \kappa < 1 \ ,
\end{eqnarray}
\begin{eqnarray}
\label{2}
    \langle t\rangle_N = \frac{1}{2}N^2 \ , \quad
    \langle t^2\rangle_N = \frac{5}{12}N^4 \ ,
    \quad\quad \kappa = 1 \ ,
\end{eqnarray}
\begin{equation}
\label{interesting2}
    \langle t\rangle_N = \frac{\kappa^{N+1}}{(\kappa-1)^2} \ ,
    \quad
    \langle t^2\rangle_N = \frac{2\kappa^{2(N+1)}}{(\kappa-1)^4} \ ,
    \quad\quad \kappa > 1 \ .
\end{equation}
\end{subequations}
These formulas already allow us to make a prediction concerning
the shape of the trapping time distribution. In particular,
comparing Eqs.~(\ref{2}) and (\ref{interesting2}) with
Eq.~(\ref{1}), we conclude that a drastic difference must exist
between the two cases $\kappa < 1$ (bias towards core) and $\kappa
\ge 1$ (no bias or bias away from core), with respect to the shape
of the trapping time distribution $R_N$. Indeed, calculating the
standard deviation $ \sigma_N = \sqrt{\langle t^2\rangle_N -
\langle t\rangle_N^2}$, we obtain
\begin{subequations}
\label{SD}
\begin{eqnarray}
    \sigma_N = \frac{N^{1/2}}{1-\kappa} \ ,
    \quad\quad \kappa < 1 \ ,
\end{eqnarray}
\begin{eqnarray}
    \sigma_N = \frac{N^2}{\sqrt 6} \ ,
    \quad\quad\quad \kappa = 1 \ ,
\end{eqnarray}
\begin{equation}
    \sigma_N = \frac{\kappa^{N+1}}{(1-\kappa)^2} \ ,
    \quad\quad \kappa > 1 \ .
\end{equation}
\end{subequations}
As is seen, for large $N$, $\sigma_N \ll \langle t\rangle_N$ for
$\kappa < 1$, i.e., the distribution $R_N$ is narrow in the sense
that its standard deviation is much smaller than its mean. By
contrast, $\sigma_N \sim \langle t\rangle_N$ if $\kappa \ge 1$,
which means that $R_N$ is a broad distribution in this sense.

\subsection{Asymptotic behavior, $N \gg 1$}
\label{asymptotics}

From Eqs.~(\ref{limit_coeff}) it follows that the characteristic
time of trapping $\langle t \rangle_N$ in dendrimers of higher
number of generation is long on the scale of the ``effective"
hopping times, which are $(1-\kappa)^{-1}$, $(\kappa-1)^{-1}$, and
$2$ [or $(k_1-k_2)^{-1}$, $(k_2-k_1)^{-1}$, and $(2k_1)^{-1}$ in
dimensional units] for dendrimers with a total bias towards the
trap ($\kappa < 1$), towards the periphery ($\kappa > 1$), and no
bias ($\kappa = 1$), respectively. Within the Laplace domain, this
means that the dominant region of the parameter $s$, being of the
order of $\langle t \rangle_N^{-1}$,  is, respectively, small
compared to $1-\kappa$, $\kappa-1$, and $2$ for these three
different situations. This allows us to significantly simplify the
expression Eq.~(\ref{closedRN}) for $\tilde{R}_N$.

\subsubsection*{1. Total bias towards the trap, $\kappa < 1$}
\label{kappasmallerthanunity}

We start analyzing the case of a total bias towards the trap
($\kappa < 1$). Using a Taylor expansion of Eq.~(\ref{closedRN})
with respect to $s/(1-\kappa)$, we obtain
\begin{equation}
\label{simple11}
    \frac{1}{\tilde{R}_N} = \frac{s\kappa^{N+1}}{(1-\kappa)^2}
    \left(1 - \frac{s}{1-\kappa} \right)^N
    + \left(1 + \frac{s}{1-\kappa} \right)^N \ .
\end{equation}
Since $\kappa < 1$ and $N \gg 1$, the first term on the left-hand
side can be neglected as compared to the second one, thus
providing us with a very simple expression for $\tilde{R}_N$
\begin{equation}
\label{simple12}
    \tilde{R}_N = \left(1 + \frac{s}{1-\kappa} \right)^{-N} \ ,
\end{equation}
which can be easily transformed back to the time domain. The result
reads
\begin{equation}
\label{ksmall}
    R_N = \frac{(1-\kappa)^N}{(N-1)!}\> \> t^{N-1} \>\> e^{-(1-\kappa)t}
    \ .
\end{equation}
We stress that this result is exact in the limit $\kappa \to 0$
(independent of $N$). As is seen, $R_N$ is strongly nonexponential
and for $N \gg 1$ is characterized by a sharp profile, consistent
with our findings in the previous subsection. In fact, in the
limit of $s/(1-\kappa) \ll 1$ and $N \to \infty$ we can
approximately write $\tilde{R}_N = \exp[-s\langle t \rangle_N]$,
where $\langle t \rangle_N = N/(1-\kappa)$ is the mean trapping
time for $\kappa < 1$. The time-domain behavior, which corresponds
to this Laplace transform, is $R_N = \delta (t - \langle t
\rangle_N)$. The nonexponentiality found is in fact a
characteristic property of the trapping in dendrimers with a total
bias towards the trap ($\kappa < 1$), independent of the number of
generations $N$.

In order to illustrate how the approximate expression
Eq.~(\ref{ksmall}) fits the exact result obtained by numerically
integrating Eqs.~(\ref{DN}), we plotted in
Figs.~\ref{fig:nonexponential2} and \ref{fig:nonexponential1} the
distribution $R_N$ calculated for $\kappa = 1/5$ and $\kappa = 1/2$
at different number of generations $N$. Here the solid lines are
the exact solutions and the dashed lines correspond to the
approximation Eq.~(\ref{ksmall}). From these plots we conclude that
for $\kappa = 1/5$, Eq.~(\ref{ksmall}) works well, even for
dendrimers with only four generations. On the other hand, for
$\kappa = 1/2$ the deviation of Eq.~(\ref{ksmall}) from the exact
solution gets larger. These figures also clearly demonstrate the
tendency of the trapping time distribution for large dendrimers to
tend towards a delta-function at the mean-first passage time.

\subsubsection*{2. Zero total bias, $\kappa = 1$}
\label{kappaequaltozero}

At zero total bias, i.e., when the energetic bias and geometrical
one compensate each other ($\kappa = 1$), the problem we are
dealing with is equivalent with the classical one-dimensional
diffusion problem on a finite segment with absorbing and reflecting
boundary conditions at $x = 0$ and $x = N$, respectively, and an
initial condition corresponding to the creation of a diffusing
object at $x = N$. The Laplace transform $\tilde{R}_N$, derived
from Eq.~(\ref{closedRN}) in the limit of $s \ll 2$, reads
\begin{equation}
\label{simple31}
        \tilde{R}_N = \frac{1}{\big(1 + \sqrt{s}\>\big)^N
        + \big(1 - \sqrt{s}\>\big)^N} \approx
        \frac{2}{\cosh\big(\sqrt{2s\langle t \rangle_N}\>\big)}\ ,
\end{equation}
where $\langle t \rangle_N = N^2/2$ is the mean trapping time in
the diffusive regime of the random walk, and we used the fact that
$(1 \pm \sqrt{s})^N \approx e^{\pm N\sqrt{s}}$ for $s \ll 1$ and $N
\gg 1$. In the time domain we then obtain
\begin{eqnarray}
\label{simple32}
        R_N = \frac{\pi}{\langle t \rangle_N}\sum_{n=0}^\infty
        (-1)^n \left(n+\frac{1}{2}\right)\>\>
        \exp\left[-\frac{\pi^2(n+1/2)^2 t}{2\langle t \rangle_N}\right] \ .
\end{eqnarray}

In Fig.~\ref{fig:exponential} (upper panel), we depicted $R_N(t)$
calculated by numertically integrating Eqs.~(\ref{DN}) for
dendrimers of different numbers of generations with $\kappa=1$.
First, we note that the curves obtained for $N = 4$ and $N = 8$
are almost identical. Second, the dotted curve in the plot,
corresponding to $N = 8$, coincides in fact with the limiting
curve; it is not changed by further increasing $N$. The decaying
part of this curve is nicely fitted by the first term of the
series Eq.~(\ref{simple32}), i.e., by $\exp(-\pi^2t/8\langle t
\rangle_N)$. Thus, we conclude that already for small $N$, the
approximate result Eq.~(\ref{simple32}) gives a good fit to the
exact result, and the tail of $R_N(t)$ is described by a single
exponential.

\subsubsection*{3. Total bias towards the periphery, $\kappa > 1$}
\label{kappalargerthanunity}

We proceed similarly to the above in the case of a total bias
towards the dendrimer periphery ($\kappa > 1$). Assuming now in
Eq.~(\ref{closedRN}) that $s/(\kappa-1) \ll 1$, one finds
\begin{eqnarray}
\label{simple21}
    \frac{1}{\tilde{R}_N} = \frac{s\kappa^{N+1}}{(\kappa-1)^2}
    \left(1 + \frac{s}{\kappa-1} \right)^N
    + \left(1 - \frac{s}{\kappa-1} \right)^N \ .
\end{eqnarray}
We recognize here $\langle t \rangle_N=\kappa^{N+1}/(\kappa-1)^2$
as the mean trapping time for $\kappa > 1$ [see
Eq.~(\ref{interesting2})]. As the relevant region of the Laplace
parameter is determined by $s\langle t \rangle_N \sim 1$, while
$\langle t \rangle_N \gg N/(k-1)$ at $\kappa > 1$, the terms
$s/(k-1)$ in the parentheses of Eq.~(\ref{simple21}) are
negligible. Upon this simplification, $\tilde{R}_N$ takes the form
\begin{eqnarray}
\label{simple22}
    \tilde{R}_N = \frac{1}{s\langle t \rangle_N + 1} \ ,
\end{eqnarray}
which, converted to the time domain, corresponds to the
exponential behavior
\begin{eqnarray}
\label{exponential} R_N\approx \frac{1}{\langle t\rangle_N}\>\>
        \exp \left[- \frac{t}{\langle t\rangle_N} \right] \ .
\end{eqnarray}
Figure~\ref{fig:exponential} (lower panel) illustrates this
finding. All curves presented in this panel were obtained by
numerically integrating Eqs.~(\ref{DN}). As is seen, all curves
are close to each other, including the one for $N = 2$. The dotted
curve is fitted very well by Eq.~(\ref{exponential}). From this we
conclude that the approximation Eq.~(\ref{exponential}) works
perfectly for any number of generations.

The difference in the behavior of $R_N$ (exponential or
nonexponential) for different signs of the total bias in principle
may be used to experimentally probe for the direction of the
energetic bias in a dendrimer. Indeed, the kinetics of the
fluorescence intensity $I(t)$ is proportional to $\psi =
e^{-\gamma t} \left[ 1 - \int_0^t R(t^\prime)dt^\prime \right]$,
i.e., is determined by two decay channels: the $\gamma$ processes
and the trapping at the core. Having found the former from, for
instance, the early-time decay of the intensity, we can then
extract information about the direction of the energetic bias by
measuring $I(t)$.

\section{Efficiency of trapping}
\label{ET}

In this section, we turn to analyzing the trapping efficiency
$\varepsilon_N = \tilde{R}_N(s)|_{s=\gamma}$ (cf.~Eq.~\ref{q}),
the exact expression for which follows directly from
Eq.~(\ref{closedRN}):
\begin{eqnarray}
\label{efficiency_exact} \frac{2^{N+1}}{\varepsilon_N} &=& \left(
1+\frac{1-\kappa+\gamma}{\sqrt{(1+\kappa+\gamma)^2-4\kappa}}
\right)
\left(1+\kappa+\gamma+\sqrt{(1+\kappa+\gamma)^2-4\kappa}\right)^N
\nonumber\\
\nonumber\\
&+& \left(
1-\frac{1-\kappa+\gamma}{\sqrt{(1+\kappa+\gamma)^2-4\kappa}}
\right)
\left(1+\kappa+\gamma-\sqrt{(1+\kappa+\gamma)^2-4\kappa}\right)^N.
\end{eqnarray}
From this result, one easily generates plots for any set of
variables $\kappa$, $\gamma$, and $N$. In particular, we present
in Fig.~\ref{fig:eff} the results for the trapping efficiency
according to Eq.~(\ref{efficiency_exact}), for dendrimers of $N=1$
to $N=10$ generations and different directions of the total bias,
setting $\gamma = 0.01$ (in units of $k_1$). The general trends
displayed in this figure are easily understood. For small
dendrimers, the efficiency is close to unity, unless a strong bias
towards the periphery is combined with a fast decay. For larger
dendrimers, the efficiency decreases due to an increased chance of
decay before reaching the core. This effect grows when increasing
the number of generations. However, a more detailed physical
interpretation of Eq.~(\ref{efficiency_exact}) for larger $N$
values requires a deeper analysis. In the following, we will focus
on this large-$N$ region. It is then natural to limit ourselves to
a decay rate $\gamma$ that is small compared to the ``effective
hopping rates" $1-\kappa$, $\kappa-1$, and $2$, as we also assumed
with respect to the Laplace parameter $s$ in our analysis of the
trapping time distribution [see Sec.~\ref{asymptotics}].
Otherwise, the trapping efficiency will be low even for a
dendrimer of a small number of generations. Hereafter, we impose
this condition, which allows us to directly use
Eqs.~(\ref{simple12}), (\ref{simple31}), and (\ref{simple22}),
replacing $s$ by $\gamma$.

\subsubsection*{1. Total bias towards the trap, $\kappa < 1$}
\label{kappasmallerthanunityET}

We first consider the case of a total bias towards the trap
($\kappa < 1$). The corresponding expression for $\varepsilon_N$ is
\begin{equation}
\label{kappasmall}
    \varepsilon_N = \left(1 + \frac{\gamma}{1-\kappa}\right)^{-N}
    = \exp\left(- \frac{\gamma N}{1-\kappa}\right)
    = \exp\left[- \gamma \langle t \rangle_N\right] \ .
\end{equation}
As is seen from this equation, the only parameter that determines
the trapping is $\gamma \langle t \rangle_N = \gamma N/(1-\kappa)$.
The interplay of trapping in the absence of excitation decay and
the excitation decay itself determines the trapping efficiency: the
latter is high (close to unity) for $\gamma\langle t\rangle_N \ll
1$, decreasing linearly with $\gamma \langle t \rangle_N$, and
exponentially small in the opposite limit, $\gamma\langle
t\rangle_N \gg 1$. We note that Eq.~(\ref{kappasmall}) implies
that in the large-$N$ limit with $\kappa < 1$, $\langle
\exp[-\gamma t] \rangle=\exp[-\gamma \langle t \rangle]$, which is
due to the fact that under these conditions the trapping time
distribution tends to a delta function at the mean trapping time,
as we have found in Sec.~\ref{kappasmallerthanunity}.

Using the definitions Eqs.~(\ref{tau}) and (\ref{W}), we can also
calculate the mean survival time $\tau_N$ and the effective
trapping rate $W_N$. They are given by
\begin{subequations}
\begin{equation}
\label{taukappasmall}
    \tau_N = \frac{1}{\gamma}\Big[1 - \exp\big(-\gamma \langle t
    \rangle_N\big)\Big] \ ,
\end{equation}
\begin{equation}
\label{Wkappasmall}
    W_N = \frac{\gamma}{\exp\big(\gamma \langle t \rangle_N\big) - 1} \ .
\end{equation}
\end{subequations}
If the decay is slow on the scale of the mean trapping time
($\gamma\langle t\rangle_N \ll 1$), the survival time $\tau_N$
coincides with the mean trapping time $\langle t \rangle_N$, and
$W_N$ is just the inverse value $\langle t \rangle_N^{-1}$. In the
opposite limit ($\gamma\langle t\rangle_N \gg 1$), we get $\tau_N
= \gamma^{-1}$ (because the excitation cannot reach the trap
within its lifetime) and $W_N = \gamma \exp[-\gamma \langle t
\rangle_N]$. Note that the range of variation of $\tau_N$ is
always from $\langle t \rangle_N$ to $\gamma^{-1}$ upon increasing
the driving parameter $\gamma\langle t\rangle_N$ from zero to
values large compared to unity. This is the general behavior of the
survival time $\tau_N$, independent of the direction of the total
bias.

\subsubsection*{2. Zero total bias, $\kappa = 1$}
\label{kappaequaltozeroET}

In the diffusive regime ($\kappa = 1$), according to
Eq.~(\ref{simple31}),
\begin{equation}
\label{kappa1}
    \varepsilon_N  =
    \frac{1}{\cosh \sqrt{2\gamma\langle t \rangle_N}} \ ,
\end{equation}
and consequently
\begin{eqnarray}
\label{Wkappasma2l}
    W_N = \frac{\gamma}{2\sinh^2\sqrt{\gamma \langle t \rangle_N/2}}
    \ ,
\end{eqnarray}
where now $\langle t \rangle_N = N^2/2$ is the trapping time for
$\kappa = 1$. In the limit of slow decay on the time scale of
trapping ($\gamma\langle t \rangle_N \ll 1$) one obtains
\begin{equation}
\label{kappa1a}
    \varepsilon_N = \frac{1}{1 + \gamma\langle t \rangle_N}
    = \frac{1}{1 +  \gamma N^2/2} \ ,
\end{equation}
i.e., the trapping efficiency is close to unity, as expected. In
Eq.~\ref{kappa1a} we kept the ``small" term in the denominator,
because this expression works well even if $\gamma\langle t
\rangle_N$ is slightly larger than unity. This is due to the fact
that the Taylor expansion of the hyperbolic cosine only contains
even powers of its argument.

If $\gamma\langle t \rangle_N$ gets larger than unity, another
regime of trapping comes into play:
\begin{equation}
\label{kappa1b}
    \varepsilon_N =
    2\exp{\Big[-\sqrt{2\gamma\langle t \rangle_N}\Big]}
    = \exp\big(-\sqrt{\gamma}N\big) \ .
\end{equation}
As is seen, asymptotically $\varepsilon_N$ decreases exponentially,
with an exponent proportional to $N$ (and not $N^2$). This thus
resembles the behavior of trapping in the case of a large bias
towards the trap (cf.~Eq.~(\ref{kappasmall}). The pre-factor
$\sqrt{\gamma}$ in Eq.~(\ref{kappa1b}), however, is larger than in
the case of a bias towards to the trap ($\gamma$). This is not
surprising, because in the diffusive regime, the excitation makes
steps towards the periphery that slow down the process of reaching
the trap, allowing for a larger effect of excitation decay before
trapping may occur.

Finally, the effective trapping rate $W_N$ starts from the value
$\langle t \rangle_N^{-1}$ in the limit of slow decay
($\gamma\langle t \rangle_N \ll 1$) and reveals the same behavior
as $\varepsilon_N$ for $\gamma\langle t \rangle_N \gg 1$.

\subsubsection*{3. Total bias towards the periphery, $\kappa > 1$}
\label{kappalargerthanunityET}

In a certain sense, this is the simplest case. For $s = \gamma$,
Eq.~(\ref{simple22}) yields for the efficiency of trapping
\begin{equation}
\label{kappalargeET}
    \varepsilon_N = \frac{1}{1 + \gamma\langle t \rangle_N}
\end{equation}
for any ratio of $\langle t \rangle_N = \kappa^{N+1}/(\kappa-1)^2$
and $\gamma^{-1}$. Consequently, the relationship $W_N = \langle t
\rangle_N^{-1}$ holds in general.

\section{Conclusions}
\label{concl}

\begin{figure}[ht]
  \includegraphics[width=12cm]{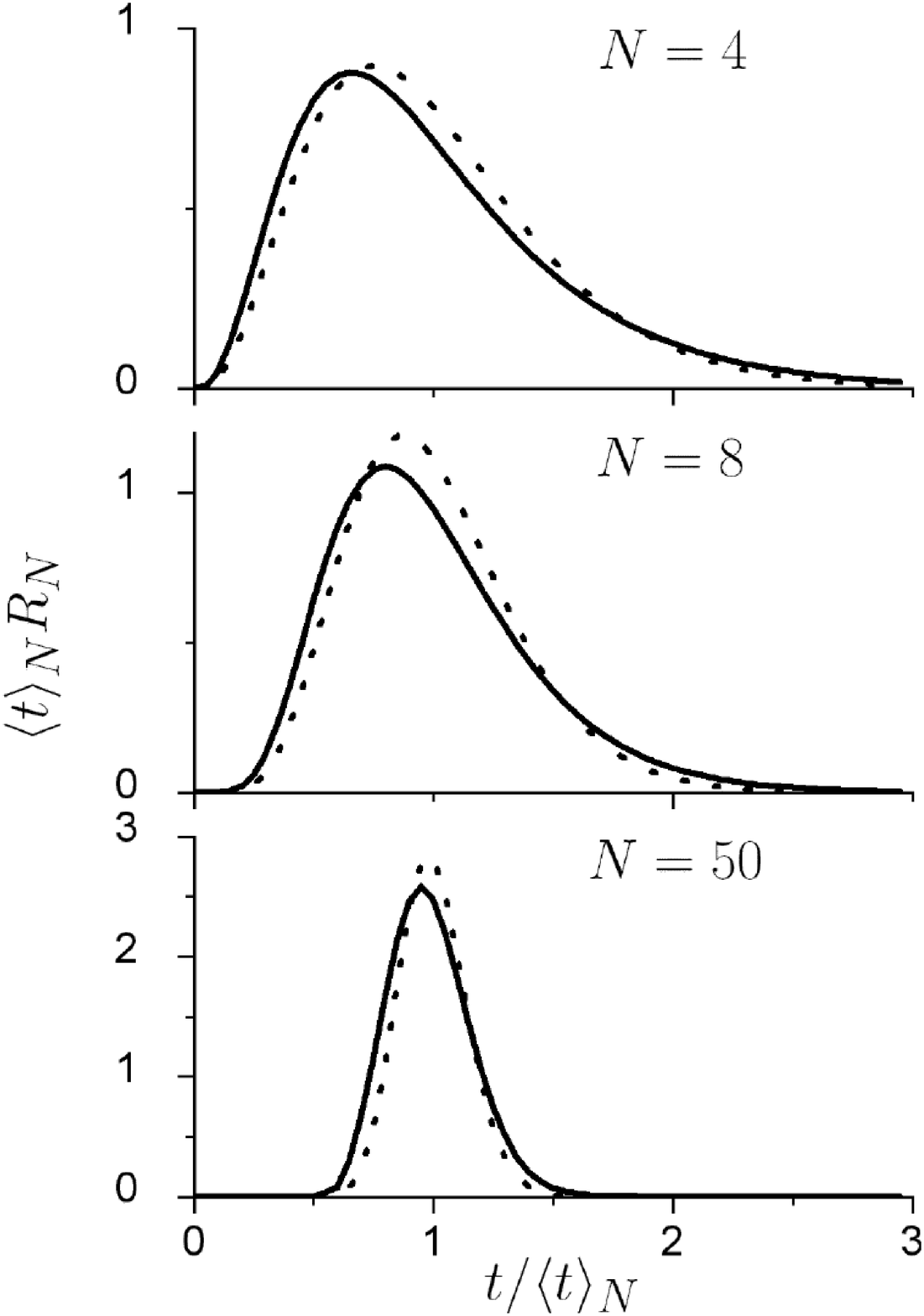}
  \caption{
  Plots of the trapping time distribution $R_N(t)$ for dendrimers
  of different number of generations $N$ with a total bias towards the
  trap ($\kappa = 1/5$).  In all plots, the solid curves represent
  the exact solution for $R_N(t)$ obtained from numerically solving
  Eqs.~(\ref{DN}), while the dashed curves correspond to the
  approximate expression Eq.~(\ref{ksmall}).
  }
\label{fig:nonexponential2}
\end{figure}

\begin{figure}[ht]
  \includegraphics[width=12cm]{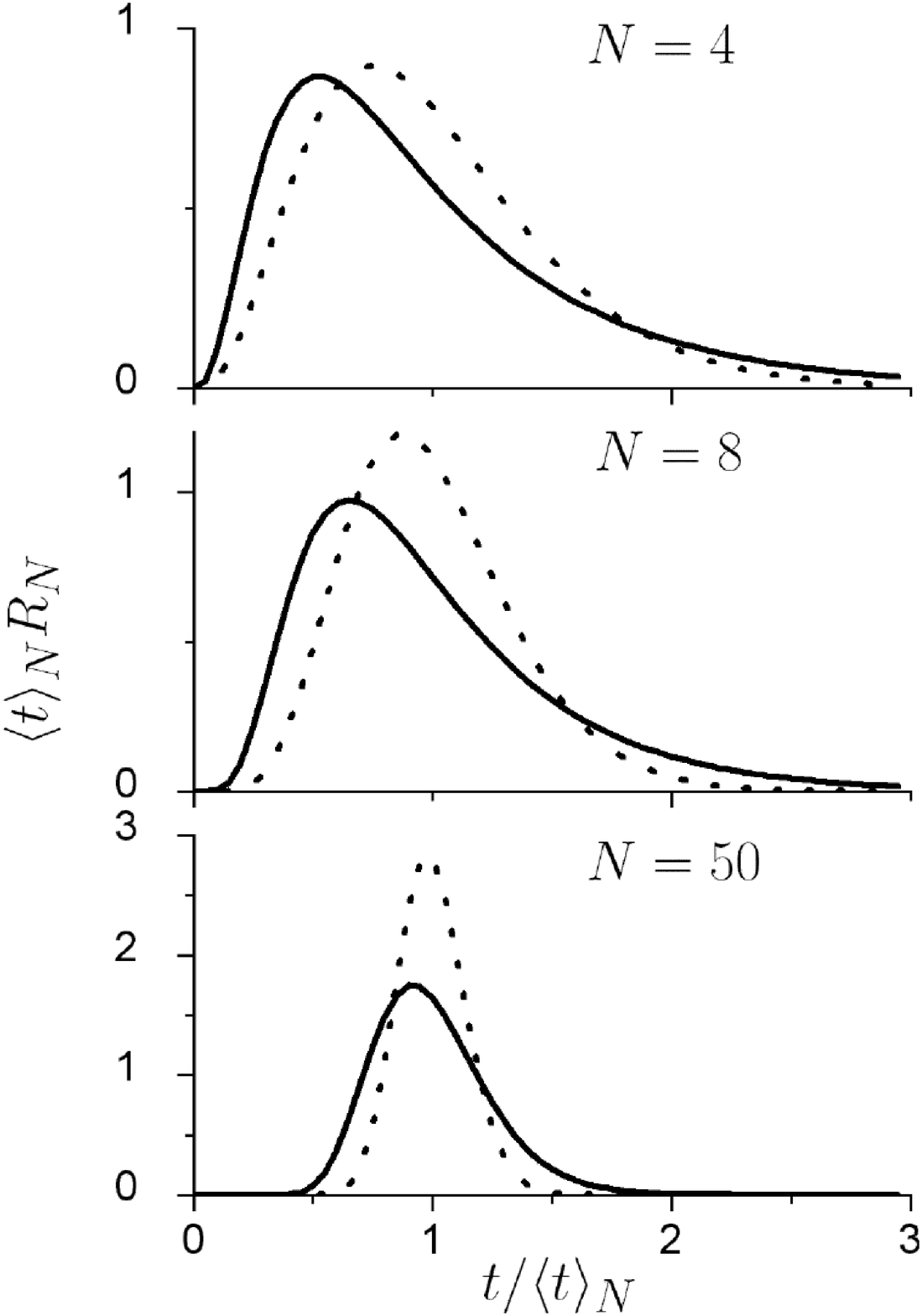}
  \caption{
  Same as in Fig.~\ref{fig:nonexponential2}, using $\kappa = 1/2$.
  }
\label{fig:nonexponential1}
\end{figure}

\begin{figure}[ht]
  \includegraphics[width=12cm]{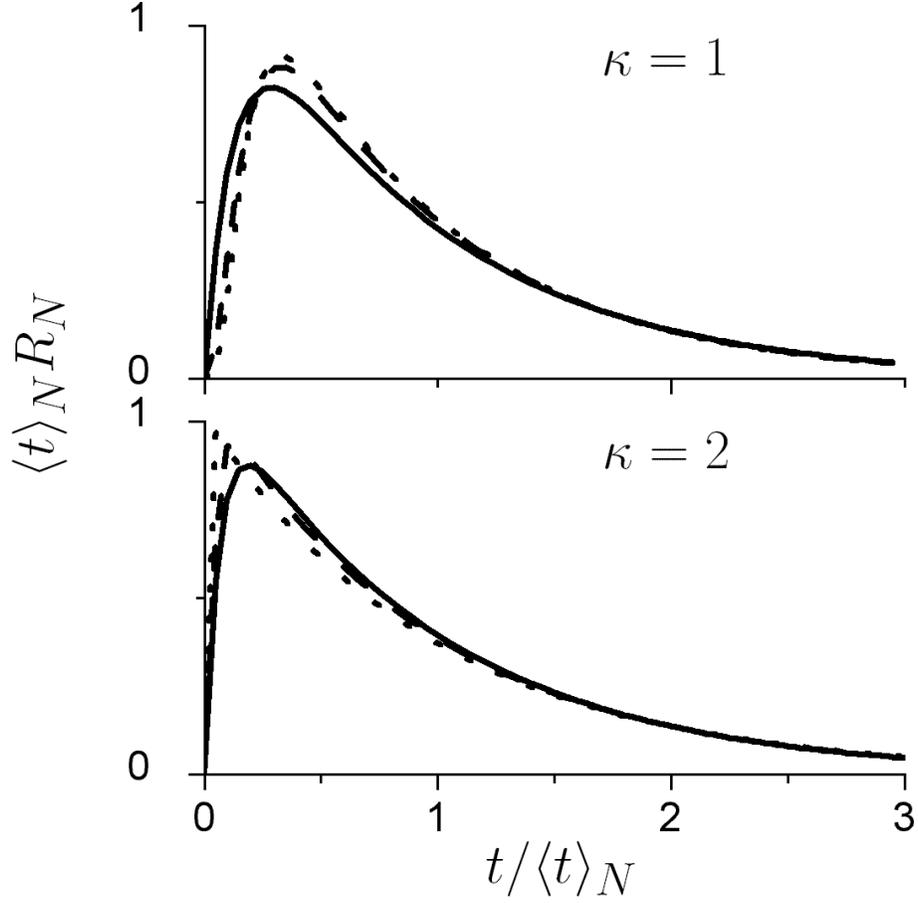}
  \caption{
  Plots of the trapping time distribution $R_N(t)$ for dendrimers
  of different number of generations $N$ in the diffusive regime
  of hopping ($\kappa = 1$) and for a total bias towards the
  periphery ($\kappa = 2$). All curves were obtained by numerically solving
  Eqs.~(\ref{DN}). In both panels, the solid curves
  correspond to $N = 2$, the dashed curves to $N = 4$, and the dotted
  curves to $N = 8$. The decaying part of the latter curve in the upper panel
  coincides with $\exp[-\pi^2t/8\langle t \rangle_N]$.
  }
\label{fig:exponential}
\end{figure}

\begin{figure}[ht]
  \includegraphics[width=12cm]{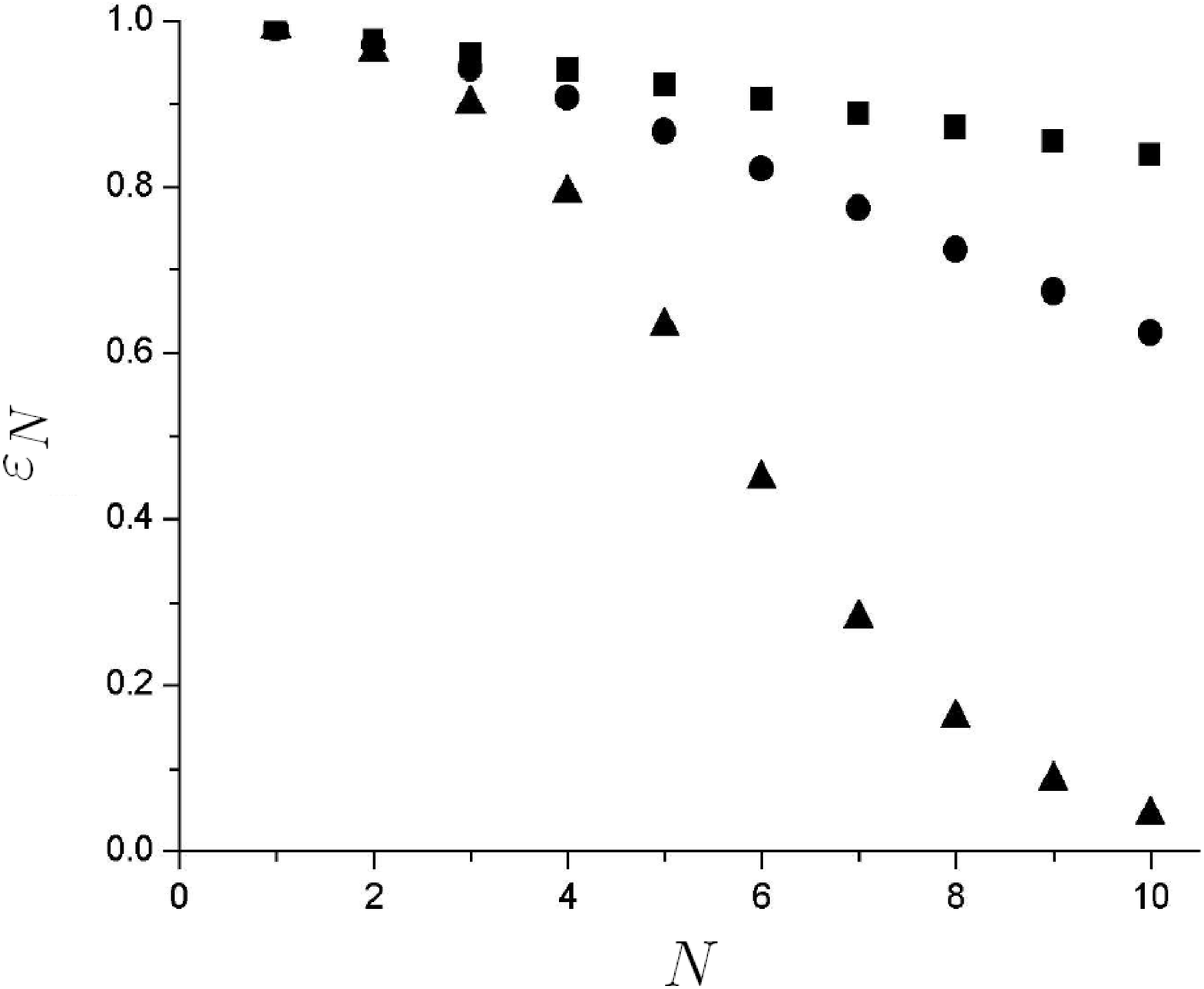}
  \caption{
  \label{fig:eff}
  Plots of the trapping efficiency $\varepsilon_N$ as a function of
  number of generations $N$ calculated for $\gamma = 0.01$ and different
  values for $\kappa$. The squares correspond to $\kappa = 1/2$ (total bias towards
  the trap), circles correspond to $\kappa = 1$ (no total bias, diffusive regime
  of hopping), and the triangles correspond to $\kappa = 2$ (total bias towards the
  periphery).
  }
\end{figure}

In this paper we theoretically studied the trapping of excitations
in dendritic systems in the presence of (radiative and
nonradiative) excitation decay when moving to the trap at the
dendrimer's center. We derived an exact expression for the Laplace
transform of the trapping time distribution for a dendrimer of any
number of generations. This expression was then used to analyze
the general properties of pure trapping (in the absence of decay),
focusing on dendrimers of a large number of generations. We found
that the general nature of this distribution is governed by the
total (geometrical and energetic) bias towards the trap. In the
presence of a bias towards the trap, the trapping time
distribution is narrow, in the sense that its standard deviation
is small compared to its mean. The shape of the distribution is
strongly nonexponential. Oppositely, in the presence of a bias
towards the dendrimer's periphery, the trapping time distribution
is broad (its standard deviation is of the order of its mean), and
its shape is essentially exponential with an exponent equal to the
mean trapping time. The strong difference between both regimes is
nicely illustrated by comparing Figs.~\ref{fig:nonexponential2} and
\ref{fig:exponential}. As the fluorescence kinetics is proportional
to the trapping time distribution, the nature of the fluorescence
decay (exponential or nonexponential) may in practice be used to
distinguish the direction of the energetic bias in dendrimers. We
also note that, although in the analysis of the trapping time
distribution we used the limit of a large number of generations,
it appears from comparison to numerically exact results that many
of the analytical expressions hold even for small dendrimers, with
just a few generations.

The trapping efficiency $\varepsilon$ was found to depend on the
ratio of the decay time ($\gamma^{-1}$) and the trapping time in
the absence of decay ($\langle t \rangle$). The product $x = \gamma
\langle t \rangle$ turns out to be the only essential parameter
that governs the trapping in the presence of excitation decay. For
dendrimers with a total bias towards the trap ($\kappa <1$), the
efficiency of trapping depends exponentially on this parameter
within its entire range of values: $\varepsilon = e^{-x}$. In the
diffusive regime of trapping, when the geometrical and energetic
bias compensate each other ($\kappa =1$), $\varepsilon = 1-x$ for
$x < 1$, while for $x > 1$ this behavior changes to a
stretched-exponential one, $\varepsilon \sim e^{-\sqrt{x}}$. For
dendrimers with a total bias towards the periphery ($\kappa >1$),
the trapping efficiency $\varepsilon = (1+x)^{-1}$, independent of
$x$.

We finally notice that in practice the various regimes with regards
to the total bias parameter $\kappa$ distinguished by us, may be
probed experimentally in one and the same dendritic system by
varying the temperature. In particular, $\kappa$ tends from zero
at low temperature to $z-1$ at high temperature.

\clearpage \newpage

\end{document}